\documentclass[twocolumn,amsmath,showkeys,prb]{revtex4-1}

\usepackage{dcolumn}
\usepackage{bm}
\usepackage[T1]{fontenc}
\usepackage{amsmath}
\usepackage{graphicx}
\usepackage{tabularx}

\usepackage{hyperref}
\hypersetup{colorlinks=true, linkcolor=blue, citecolor=red, urlcolor=magenta, pdftitle={SIA paper}, pdfauthor={Eric Bousquet}}

\begin{document}
\title{Non collinear magnetism and single ion anisotropy in multiferroic perovskites}
\author{Carlo Weingart$^1$, Nicola Spaldin$^1$ and Eric Bousquet$^{1,2}$}
\affiliation{$^1$ Department of Materials, ETH Zurich, Wolfgang-Pauli-Strasse 27, CH-8093 Zurich, Switzerland}
\affiliation{$^2$Physique Th\'eorique des Mat\'eriaux, Universit\'e de Li\`ege, B-4000 Sart Tilman, Belgium}

\begin{abstract}
The link between the crystal distortions of the perovskite structure and the magnetic exchange interaction, the single-ion anisotropy (SIA) and the Dzyaloshinsky-Moriya (DM) interaction are investigated by means of density-functional calculations.
Using BiFeO$_3$ and LaFeO$_3$ as model systems, we quantify the relationship between the oxygen octahedra rotations, the ferroelectricity and the weak ferromagnetism (wFM).
We recover the fact that the wFM is due to the DM interaction induced by the oxygen octahedra rotations.
We find a simple relationship between the wFM, the oxygen rotation amplitude and the ratio between the DM vector and the exchange parameter such as the wFM increases with the oxygen octahedra rotation when the SIA does not compete with the DM forces induced on the spins.
Unexpectedly, we also find that, in spite of the $d^5$ electronic configuration of Fe$^{3+}$, the SIA is very large in some structures and is surprisingly strongly sensitive to the chemistry of the $A$-site cation of the $A$BO$_3$ perovskite.
In the ground $R3c$ state phase we show that the SIA shape induced by the ferroelectricity and the oxygen octahedra rotations are in competition such as it is possible to tune the wFM ''on`` and ''off" through the relative size of the two types of distortion.
\end{abstract}

\maketitle

\section{Introduction}
During the last ten years, there has been a huge increase of interest in developing magnetoelectric multiferroic materials. 
In such materials, ferroelectric and magnetic ordering coexist together and can be coupled such that the magnetization is affected by an electric field and the polarization by a magnetic field. 
The reason for this interest is related to impact potential in technological applications in transducers, attenuators, filters, information storage and spintronics \cite{spaldin2005,fiebig2005,bibes2008,tokunaga2009,feng2010}.
The ideal magnetoelectric multiferroic would be a compound in which a large spontaneous polarization were coupled with a large magnetization so that flipping the former could flip the latter and vice-versa.
However, such an ideal compound is not known today.
A particularly promising direction is to switch magnetization by 180$^\circ$ using an electric field in materials exhibiting weak ferromagnetism (wFM) \cite{fennie2008, bousquet2008}.
In weak ferromagnets, the magnetization is small, but it has been proposed theoretically that in the presence of a ferroelectric polarization one can switch the wFM by reversing the polarization \cite{fennie2008, benedek2011, bousquet2008, rondinelli2012}.

Following this pathway for magnetoelectric control of wFM by electric polarization, it is crucial to understand the underlying mechanisms that link the wFM to the crystal distortions. 
Dzyaloshinsky and Moriya (DM) showed that spin-orbit interaction (SOI) mediates a spin-spin coupling of the form $D.S_1\times S_2$ (the so-called DM interaction) that is usually responsible for wFM \cite{dzyaloshinsky1958, moriya1960}. 
Moreover, Bertaut showed by symmetry considerations that under sufficiently low crystal symmetry, SOI also permits 
 the single ion anisotropy (SIA) to cause non-collinear magnetic arrangements \cite{bertaut1963}.
However, the latter interaction has been less well studied, and in most cases only the DM interaction has been considered to be responsible for the wFM.

In multiferroic perovskites, the initial para-electric cubic reference structure can be deformed by the presence of two main lattice instabilities: the antiferrodistortive (AFD) instabilities, consisting of non-polar oxygen octahedra rotations, and the ferroelectric (FE) instability, responsible for the polarisation. 
It has been recognized that spin canting is induced by the AFD distortions and thus the wFM is directly linked to the amplitude of the AFDs \cite{ederer2005a,kim2011}.
From symmetry analysis, some of such distorted structures do not allow for spin canting while some allow only the DM to be responsible for the spin canting and a few cases allow spin canting through both DM and SIA \cite{bertaut1963}.
In spite of this, in multiferroic perovskites, no systematic study has been performed to analyze the details of the coupling between spin canting and lattice distortions.

The aim of the present study is to quantify from first-principles calculations the mechanisms leading to wFM in two representative perovskites, LaFeO$_3$ (LFO) and BiFeO$_3$ (BFO). 
Both compounds show AFD distortions (of different types), and BFO also has a FE instability, whereas LFO does not.
To understand the links between these lattice distortions and the magnetism, we decompose the magnetic interaction into three main types, exchange, DM and SIA and look at how they are affected by the amplitude and combinations of the different lattice distortions.
This systematic analysis allows us to understand the coupling between structural distortions and spin canting, and propose some guidelines for the design of magnetoelectric wFMs through the SIA.

\section{Technical details}
The first-principles calculations were performed using density-functional theory (DFT) as implemented in the VASP code within the Projector Augmented Wave (PAW) method \cite{vasp1, vasp2}.
Local spin density approximation with an additional Hubbard (LSDA$+U$) was used for the exchange-correlation functional. 
The Hubbard parameter $U$ and the exchange interaction $J$ that treat the Fe $d$ electrons were set to $U$=5 eV and $J$=0 eV, values that have been shown to be optimal for Fe$^{3+}$ in LaFeO$_3$ and BiFeO$_3$ \cite{ederer2005a,hong2012,bousquet2010b}.
The wave functions were expanded in plane waves up to a cut-off energy of 500 eV. 
For integration of the Brillouin-zone of the supercell made of $2\times2\times2$ cubic perovskite units, we used a $3\times3\times3$ Monkhorst-Pack k-point mesh shifted by $\frac{1}{2}\times\frac{1}{2}\times\frac{1}{2}$. 
The convergences were tested with a cut-off energy up to 700 eV and a $6\times6\times6$ k-point mesh, with no significant change on the calculated SIA.
Spin-orbit coupling (SOC) was included in calculating both total energies and forces to incorporate the coupling between the spins and the lattice.
Symmetrization was switched off to remove any artificial constraints on the possible spin ordering.

AFD distortions are described by rotating the oxygen octahedra around the central Fe$^{3+}$ ion, where the rotation axis is one of the Cartesian axis. 
We use the modified Glazer notation \cite{glazer1972}, where the complete AFD pattern is classified by a triplet $a^{\alpha}b^{\beta}c^{\gamma}$ with $a$, $b$ and $c$ refering to the amplitude of rotation around the Cartesian $x$, $y$ and $z$-axis and the superscripts $\alpha,\beta,\gamma=\{+,-,0\}$ refer to the type of rotation, $+$ and $-$ stand for in-phase and out-of-phase respectively and $0$ for no rotation in this direction.

\begin{figure}[htbp!]
\centering 
\includegraphics[width=8cm,keepaspectratio=true]{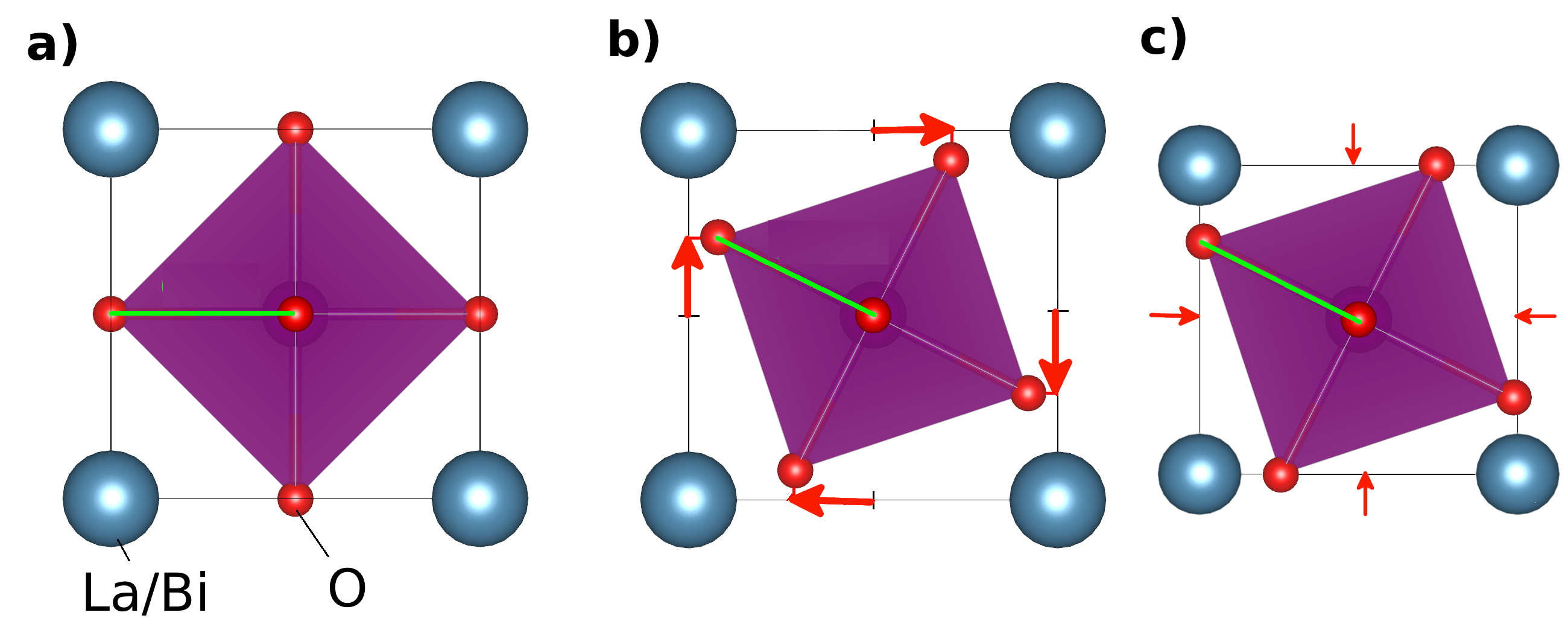}
\caption{Schematic view of the geometrical skrink of the perovskite cell in the presence of an AFD distortion along z-axis perpendicular to the paper plane.}
\label{fig:rotation}
\end{figure}

To control the amplitude of AFD rotations, we treated the Fe--O bond length as constant when freezing in the oxygen octahedra rotations, which is a good approximation with respect to the fully relaxed AFD structures. 
The rotation is sometimes thought of as the displacement of the oxygen along the edge of the cubic unit cell, which is pictured by the red arrows in Fig.\ref{fig:rotation}.b.
However, this view is valid for very small oxygen octahedra rotation only since for large amplitude of rotations, if the oxygen are kept on the edge of the cubic cell, this will strongly stretch the Fe--O bonds in the plane of rotation, a distortion that is not observed in fully relaxed structures.
Instead, if the Fe--O bond lengths are kept constant (we chose 3.90 \AA, which is close to the Fe--O bond lenght observed in the ground state of BFO\cite{palevicz2010} and LFO\cite{shivakumara2006}), an oxygen octahedra rotation is accompanied by a geometric shrinkage of the cell parameter in the plane perpendicular to the rotation axis as illustrated in Fig.\ref{fig:rotation}.b and c.

The interaction parameters, exchange $J_{ij}$, DM $D_{ij}$, and SIA $\Phi_{ii}$, in the Hamiltonian\cite{bertaut1963} 
\begin{equation}\nonumber
\mathcal{H} = -2 [J_{ij} S_i \cdot S_j + D_{ij}\cdot(S_i \times S_{j})
+ S_i \cdot \Phi_{i} \cdot S_i]
\end{equation}
were extracted from first-principles calculations.
We used a $2\times2\times2$ supercell with 8 $A$BO$_3$ formula units. 
To separate these three interactions, we used two approaches.
First, we performed artificial calculations where selected Fe$^{3+}$ ions were replaced by nonmagnetic Al$^{3+}$ ions so that only one of the three interactions was retained. 
For example, the SIA is extracted by replacing all except one of the Fe cations with neutral Al cations and then  performing constrained calculations of the direction of the remaining spin (Lagrange multiplier) to resolve the energy surface.
The data points were then fitted to extract the parameters $K_i$ on the usual expressions for the SIA \cite{skomski2008}.
By removing all the surrounding magnetic cations, we assure the vanishing of all magnetic interactions (exchange and DM) except for the SIA \cite{gonchar2000}.
We note that in this special configuration the point symmetry of the remaining Fe atom is the same as in the configuration where all the Fe atoms are present, a condition that guarantees that the crystal field splitting will be similar in the two configurations.
The exchange parameters ($J_i$) and DM vectors ($D_{ij}$) are extracted by replacing all except for two of the Fe cations with Al.
The exchange constant can then be extracted directly from the energy difference between parallel and anti-parallel arrangements of the two spins using 
\begin{equation}\nonumber
 J_{ij}       = \frac{1}{3} \sum_{a=x,y,z} \frac{1}{4S^2} (E[\hat{s}_a^i,\hat{s}_a^j] - E[\hat{s}_a^i,-\hat{s}_a^j] )
\end{equation}
where $S$ is the spin moment ($\mu_B$), $E[\hat{s_a},\hat{s_a}]$ is the total energy of a spin configuration and $\hat{s}_a$ the spin direction.
The DM vectors can be extracted by perpendicular arrangements of the two spins using
\begin{equation}\nonumber
 [D_{ij} ]_a  = \frac{1}{4S^2} (E[\hat{s}_b^i,\hat{s}^j_c] - E[\hat{s}^i_c,\hat{s}_b^j] ) \\
\end{equation}
with $a$, $b$ and $c$ being three perpendicular orientations.
This method is valid only if both magnetic ions have the same non-uniaxial SIA or if the SIA is uniaxial, which is the case in the structures analysed in the present study.
A second approach is that proposed by Xiang \emph{et al.},\cite{xiang2011} which does not have this restriction on the symmetry of the SIA, but requires twice as many calculations.
We compared the results from both methods and find simular results.

\section{Exchange interaction}

In Tab.\ref{tab:J} we report the calculated exchange parameters $J_{ij}$ extracted from our DFT calculations for LaFeO$_3$ and BiFeO$_3$ in cubic, $Pnma$, $R$\={3}c and $R3c$ structures (cell parameters and amplitude of distortions are kept fixed to the same values for LFO and BFO in each case).
The cubic structure always shows stronger exchange interactions than the distorted structures because of the large orbital overlap associated with its 180$^\circ$ Fe--O--Fe bonds.
It is interesting to see that the $J$'s for BFO and LFO are very close in all the structures, we note, however, that in BFO the exchange parameters are always smaller than in LFO due to the fact that both BFO and LFO have the same Fe$^{3+}$ magnetic cation and we are imposing the same environment.
We also remark that the $J$'s decrease with increasing the amount of either AFD or FE structural distortion.
In addition to the method described in the previous chapter, we also computed these exchange parameters with the method proposed by Xiang \cite{xiang2011} which gives similar results. 
For example we find with this second method $J_{ac}=$ 6.70 meV and 6.06 meV respectively for the $Pnma$ and $R\bar3c$ phase of LFO. 

\begin{table}[htbp!]
\centering
\begin{tabularx}{\columnwidth}{m{1cm} m{1.5cm} m{2.6cm} m{2.0cm} c}
\hline
\hline
 & \multicolumn{2}{c}{Structure}& J$_{ac}$	& J$_b$ \rule[-1ex]{0pt}{3.5ex}\\
\hline
LFO & cubic& $0^00^00^0$	& 7.53			&  --	\rule[-1ex]{0pt}{3.7ex}\\
    & $Pnma$ & $7^-8^+7^-$	& 6.74 [6.70]			& 6.83	\\
    & $R\bar3c$ & $9^-9^-9^-$	& 6.17 [6.06]			&  --	\\
    & $R3c$ & $9^-9^-9^-$+1.0 FE	& 5.64			&  --	\\
BFO & cubic& $0^00^00^0$	& 7.36			&  --	\rule[-1ex]{0pt}{4.2ex}\\
    & $Pnma$ & $7^-8^+7^-$	& 6.52			& 6.68	\\
    & $R\bar3c$ & $9^-9^-9^-$	& 5.96			&  --	\\
    & $R3c$ & $9^-9^-9^-$+1.0 FE	& 5.36			&  --	\\
\hline
\end{tabularx}
\caption{Our calculated Heisenberg exchange constants (meV) for LaFeO$_3$ and BiFeO$_3$ in different crystal structures.
We used a modified Glazer notation to indicate the amplitude of the oxygen octahedra rotations (in degree) in the three directions.
The amplitude of the FE distortions in the $R3c$ structures (1.0 FE) are those given by the minimum of energy when freezing in the polar unstable mode of the corresponding paraelectric reference $R\bar3c$ structure.
The numbers in brackets were obtained with the method of Xiang \emph{et al.} \cite{xiang2011}.}
\label{tab:J}
\end{table}

Because all the exchange interaction constants are positive, the favored spin structure is antiferromagnetic G-type for all the structures. 
This is confirmed by looking at the energy differences between the $G$, $C$, $A$ and $F$ types of magnetic orders  reported in Tab.\ref{tab:magn_ordering}, where it is clear that the lowest energy state is always $G$-type AFM order.

\begin{table}[htbp!]
\centering
\begin{tabularx}{\columnwidth}{X X X X r}
\hline
\hline
Magnetic & \multicolumn{2}{c}{LFO}&  \multicolumn{2}{c}{BFO} \rule[-1ex]{0pt}{3.5ex}\\
ordering & cubic	&	$Pnma$		& cubic		& $Pnma$		\\
\hline				
$A$-type				& -0.12	& -0.10			& -0.11		& -0.10	\rule[-1ex]{0pt}{3.5ex}\\
$C$-type				& -0.22 & -0.19			& -0.21		& -0.19 \rule[-1ex]{0pt}{3.5ex}\\
$G$-type				& -0.31 & -0.27			& -0.30		& -0.27 \rule[-1ex]{0pt}{3.5ex}\\
\hline
\end{tabularx}
\caption{Energy difference per formula unit (in eV) of the three main antiferromagnetic orders with respect to the ferromagnetic F-type configuration.}
\label{tab:magn_ordering}
\end{table}

Next we calculate how the DM interaction is affected by the same structural distortions.

\section{DM interaction}
In this section we compare the amplitude of the DM vectors between BFO and LFO in $Pnma$, $R\bar3c$ and $R3c$ structures.

\subsection{$Pnma$ structure} \label{sec:DM-pnma}
\begin{figure}[htbp!]
\centering 
\includegraphics[width=8.3cm,keepaspectratio=true]{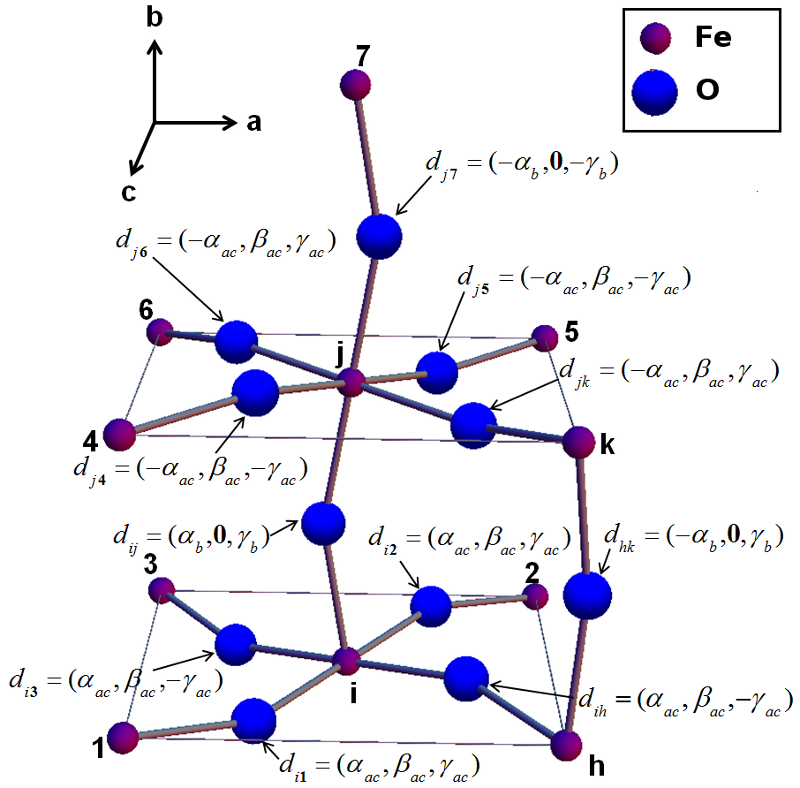}
\caption{DM vectors located at the oxygen atom between two Fe for the $Pnma$ structure.}
\label{fig:pnma_dm_abc}
\end{figure}

The symmetry elements of the orthorhombic $Pnma$ space group  and the antisymmetry of the DM interaction ($D_{ij} = - D_{ji}$) determine that the DM vectors have the form shown in Fig.\ref{fig:pnma_dm_abc} described by five parameters: $\alpha_{ac}, \beta_{ac}, \gamma_{ac}, \alpha_b , \gamma_b$.\cite{xiang2011}. 
Note that the $\mathbf{D_{ij}}$ vectors corresponding to two Fe moments are located on the intermediate oxygen ions.
The DM vector at each O site is known to be perpendicular to the mirror plane running through the associated Fe--O--Fe unit\cite{dzyaloshinsky1958}, and given by $D_{ij} \propto \hat{x}_i \times \hat{x}_j$, where $\hat{x}_k$  is the bonding vector O--Fe$_k$

\begin{figure}[htbp!]
\centering 
\includegraphics[width=8cm,keepaspectratio=true]{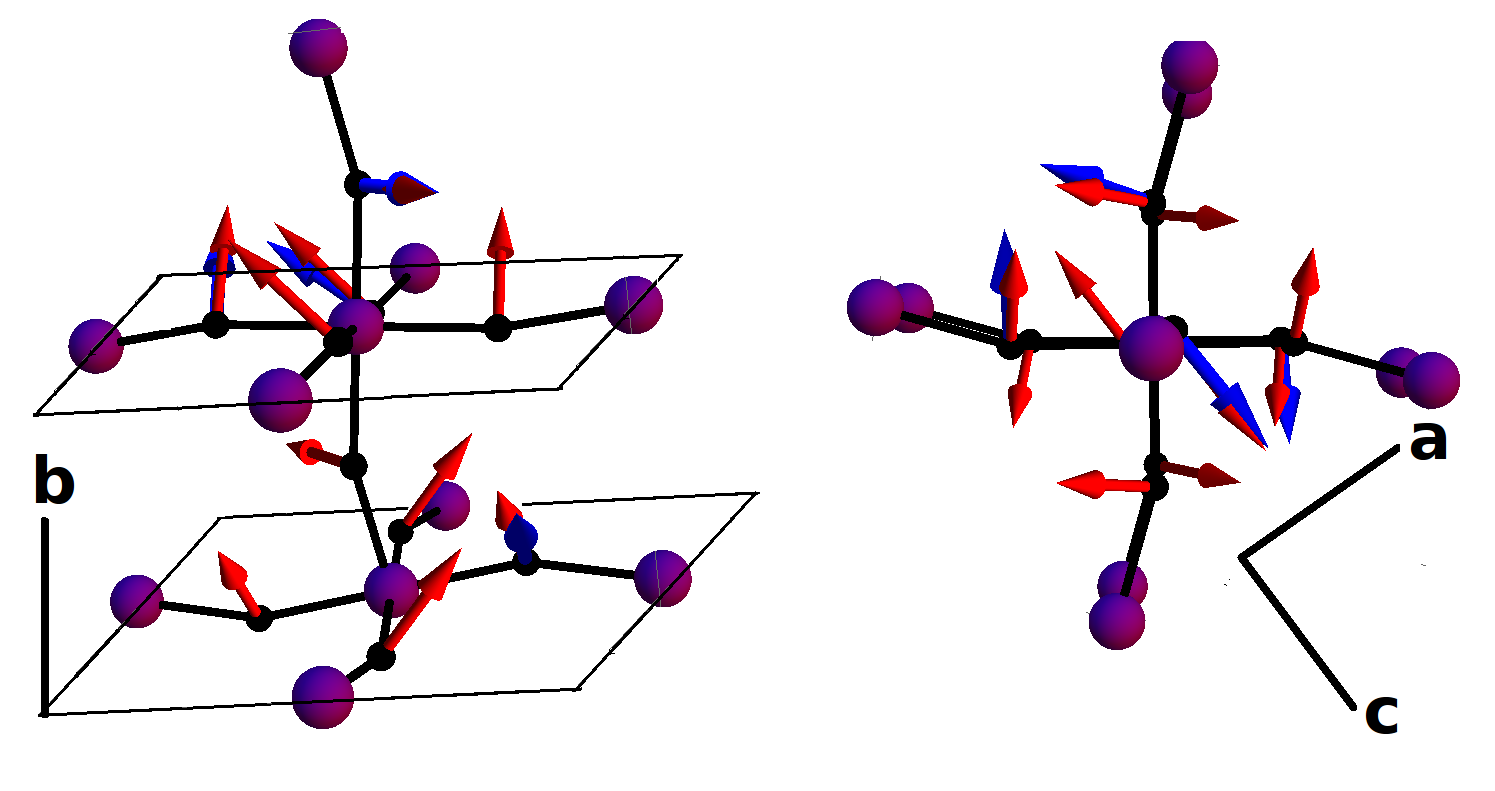}
\caption{DM vectors of two adjacent Fe ions along the \textbf{b} direction shown from two different angles. 
The directions of the  geometrically constructed  (calculated) DM vectors $D_{ij}$ are shown in red (blue).}
\label{fig:pnma_dm}
\end{figure}

In Tab.\ref{tab:pnma_dm} we report our calculated parameters of the DM vectors for both LFO and BFO.
The absolute values of the components of the total DM vector, which are the sums of the local DM vectors shown in Fig.\ref{fig:pnma_dm_abc}, are reported in Tab.\ref{tab:pnma_dm}.
The amplitude of the DM vectors are of the order of several hundred $\mu$eV. 
Comparing the values with those calculated for LFO by Kim et al.\cite{kim2011}, we find that our calculated absolute values are smaller but the ratio $D_{ac}/J_{ac}$ = 0.018 and $D_b/J_b$ = 0.017 are consistent with the values of Kim et al. who reported 0.020 and 0.021 respectively. 
One possible reason for the difference could be the different value of $U$ used (7.5 eV in the microscopic model of Ref.\onlinecite{kim2011}).
In Fig.\ref{fig:pnma_dm} we show the corresponding directions for these calculated DM vectors in LFO (blue arrows) which are in good agreement with those we expect from geometrical constructions (red arrows).

\begin{table}[htbp!]
\centering
\begin{tabularx}{\columnwidth}{X X X X X X X X r}
\hline
\hline
 & $\alpha_{ac}$  & $\beta_{ac}$ & $\gamma_{ac}$  & $\alpha_{b}$ \rule[-1ex]{0pt}{3.5ex}&
$\mathbf{\gamma_{b}}$  & $|D_x|$ & $|D_y|$ & $|D_z|$ \rule[-1ex]{0pt}{3.5ex}\\
\hline
LFO 		& 66 & 91 & 50 & 115 & 3      & 494 & 364 & 6 \rule[-1ex]{0pt}{3.5ex}\\
BFO 		& 69 & 52 & 0  & 89  & 62     & 454 & 208 & 124 \rule[-1ex]{0pt}{3.5ex}\\
LFO \cite{kim2011} 	& 99 & 127&109 & 212 & 30 & 820 & 508 & 60 \rule[-1ex]{0pt}{3.5ex}\\
\hline
\end{tabularx}
\caption{DM vector components ($\mu$eV) of LFO and BFO in the $Pnma$ phase ($7^-8^+7^-$), calculated in this work (top two rows) and in \cite{kim2011} (bottom row).}
\label{tab:pnma_dm}
\end{table}

The DM interaction between two spins $\mathbf{s_i}$ and $\mathbf{s_j}$ can be seen as a force induced by the spin $\mathbf{s_j}$ on the spin $\mathbf{s_i}$ and can be expressed as $\mathbf{f^j_i}= \mathbf{D_{ij}} \times \mathbf{s_j}$.\cite{solovyev1996}
The total force acting on the spin $\mathbf{s_i}$ due to its six neighboring spins $\mathbf{s_j}$ is thus 
$\mathbf{f_i} = \sum_{j=1}^6 \mathbf{f^{j}_{i}}$.
Using this force analysis of the DM interaction, in the $Pnma$ structure with G$_z$-type ordering ($G$-type where the spins lie along the $z$ direction), one finds two types of force having the same norm: $\mathbf{f_1} = \left(-4\beta_{ac},-2\alpha_b - 4\alpha_{ac}, 0 \right)$ and  $\mathbf{f_2} = \left(+4\beta_{ac}, -2\alpha_b - 4\alpha_{ac}, 0\right)$.  
These two types of force differ only in their orientation along the $x$ direction and show that half of the spins feel a force along $+x$ and half along $-x$.
Therefore we see that the DM interaction parameter $\beta_{ac}$ causes the $A$-type canted AFM predicted by symmetry along the $x$ direction ($A_x$).
Along the $y$ direction all the spin sites feel a force in the same direction which causes the wFM canting along the $y$ direction ($F_y$) with a strength determined by $\alpha_b$ and $\alpha_{ac}$.
Along the easy $z$ direction, no force is induced by the DM interaction.
This is in good agreement with simple symmetry considerations that show that $G_z$, $A_x$ and $F_y$ have the same symmetry transformation in the $Pnma$ perovskite structure \cite{bousquet2011b,bertaut1961,bertaut1963}.
The resulting torques are $\left(\pm364,-446, 0\right)$ $\mu$eV for LFO and $ \left(\pm208,-454, 0 \right)$ $\mu$eV for BFO.
The forces induced by the DM interaction are of the same amplitude along the $y$ direction ($F_y$ canting) in both LFO and BFO, while the resulting force along the $x$ direction ($A_x$ canting) is larger in the case of LFO than in BFO.

\subsection{DM in \textit{R\={3}c and R3c}} \label{sec:dm_r3c}

\begin{figure}[htbp!]
\centering 
\includegraphics[width=4.5cm,keepaspectratio=true]{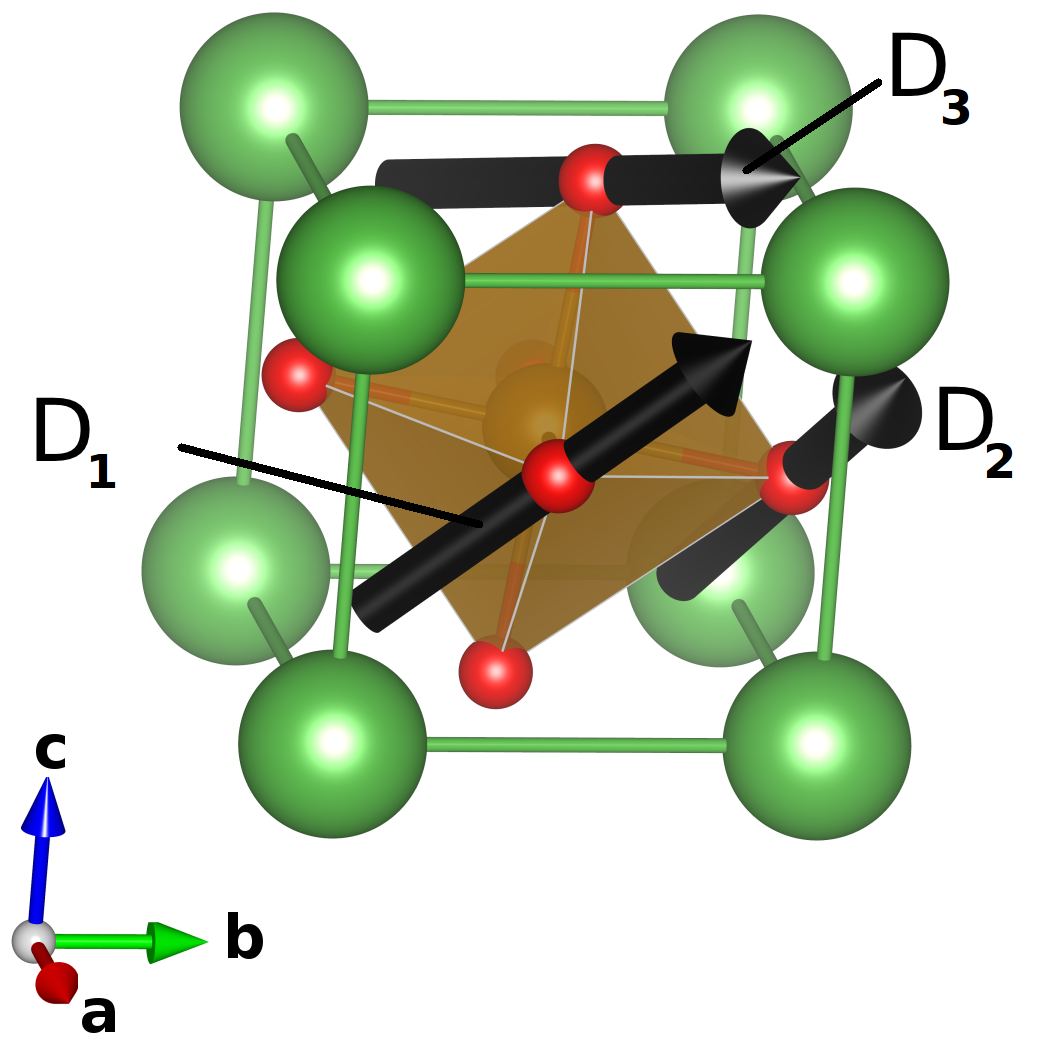}
\caption{Sketch of the orientation of the three local DM vectors in $R\bar3c$ and $R3c$ structures.}
\label{fig:r3c_dm}
\end{figure}

In $R\bar3c$ and $R3c$, the symmetry of the crystal structure requires that the total DM vector is along the [111] direction \cite{ederer2005a}. 
The direction of canting is related to the sign of $\mathbf{D}$ in such a way that the three vectors $\mathbf{D}$, $\mathbf{s_1}$ and $\mathbf{s_2}$ build up a right-handed system.	
The three local DM vectors (Fig.\ref{fig:r3c_dm}) can be fully described using only two independent parameters $\alpha$ and $\beta$: $\mathbf{D_1} = \left(\beta, \alpha, \alpha \right)$, $\mathbf{D_2} = \left(\alpha, \beta, \alpha\right)$, $\mathbf{D_3} = \left(\alpha, \alpha, \beta  \right)$.

In Tab.\ref{tab:r3c_dm} we report our calculated values of the $\alpha$ and $\beta$ parameters and the total DM vector for LFO and BFO in both $R\bar3c$ and $R3c$ phases.
$\alpha$ is always much larger than $\beta$ in LFO while in BFO they have the same amplitude.
The total magnitude of the DM vectors are however similar for LFO and BFO as well as in the two structures $R\bar3c$ and $R3c$.
The FE distortions have the tendency to reduce the DM vector but with a relatively small effect.

\begin{table}[htbp!]
\centering
\begin{tabularx}{\columnwidth}{m{1cm} X X m{1cm} c}
\hline
\hline
 &  & $\mathbf{\alpha}$  & $\mathbf{\beta}$ & $|D_x|=|D_y|=|D_z|$  \rule[-1ex]{0pt}{3.5ex}\\
\hline
$R\bar3c$ & LFO  & 92  & 12  & 196 \rule[-1ex]{0pt}{3.5ex}\\
          & BFO  & 52  & 66  & 170 \rule[-1ex]{0pt}{3.5ex}\\
$R3c $    & LFO  & 50  & 20  & 120 \rule[-1ex]{0pt}{3.5ex}\\
          & BFO  & 48  & 50  & 146 \rule[-1ex]{0pt}{3.5ex}\\
\hline
\end{tabularx}
\caption{Calculated $\alpha$ and $\beta$ parameters ($\mu$eV) and magnitude of the DM vector components ($\mu$eV) of LFO and BFO in the $R\bar3c$ and $R3c$ phases.}
\label{tab:r3c_dm}
\end{table}

As we did for the $Pnma$ phase, we can compute the forces induced on the spins due to the DM interaction.
In the case of $R\bar3c$ one finds that all the spin sublattices feel the same force $\mathbf{f} =  (-\beta -2 \alpha, -\beta -2\alpha , \beta + 2\alpha)$.
This force is perpendicular to the spins and lies in the (111) plane which causes a canting of all the spins in the (111) plane.
These results are in good agreement with the experimental data where the wFM moment is found to be in the plane perpendicular to the [111] direction.\cite{sosnowska1982,ramazanoglu2011}

In the $Pnma$ and $R\bar3c$ structures, we find that the DM interaction has the tendency to cant the spins away from the easy axis, related to the wFM or wAFM observed in these structures.
In the next section we calculate the effect of the SIA on the magnetic ground state of BFO and LFO.

\section{Single ion anisotropy} 
Finally, we calculate the behaviour of the SIA energy in the cubic and different distorted structures, focusing in particular on how the AFD and FE distortions and their combinations affect SIA.
Because the SIA has a more complex link with the crystal distortions than the DM and exchange interactions, we analyse the SIA for $a^0a^0c^+$, $a^0a^0c^-$, $a^0b^+b^+$, $a^0b^-b^-$, in addition to cubic, $Pnma$, $R\bar3c$ and $R3c$ structures.

\subsection{Cubic perovskite}
A perovskite without any oxygen octahedra rotation has the cubic space group \textit{Pm3m} \cite{glazer1972}.
For cubic crystal field splitting, the SIA can be described by 4$^{th}$ and 6$^{th}$ order terms:
\begin{equation}
 E_{SIA}^{cubic} (\alpha_i)=K_1^c(\alpha_x^2 \alpha_y^2 + \alpha_y^2 \alpha_z^2 + \alpha_z^2 \alpha_x^2) + K_2^c (\alpha_x^2 \alpha_y^2 \alpha_z^2)
\label{eq:sia_cubic}
\end{equation}
where $\alpha_i$ is the normalized projection of the spin in the $i$-direction with the constraint $\sum_{i=1}^3 \alpha_i^2 \equiv 1$ and $K_1$ and $K_2$ are the SIA parameters. 
Our calculated SIA constants obtained for LFO and BFO in the cubic structure are given in Tab.\ref{tab:sia_cubic}.

\begin{table}[htbp!]
\centering
\begin{tabularx}{\columnwidth}{X X r}
\hline
\hline
 & $K_1^c$ & $K_2^c$  \rule[-1ex]{0pt}{3.5ex} \\
\hline
LFO & -1.62  & 0.00 \rule[-1ex]{0pt}{3.5ex}  \\
BFO & -3.66 & 0.06 \rule[-1ex]{0pt}{3.5ex} \\
\hline
\end{tabularx}
\caption{Calculated SIA constants of LFO and BFO in the cubic structure fitted to Eq.\ref{eq:sia_cubic}.}
\label{tab:sia_cubic}
\end{table}

As we can see, $K_1^c$ is negative for both BFO and LFO, which indicates that the spins point in any of the three diagonal directions [111]\cite{skomski2008}.
As expected for cubic symmetry, the anisotropy energy is very small (a few $\mu$eV). 
Interestingly, it is two times larger in BFO than in LFO.
The 6$^{th}$ order anisotropy constant $K_2^c$ is close to zero for LFO and is two orders of magnitude smaller than $K_1^c$ in BFO.
This shows that, in the cubic structure, the chemistry of the A-site ion affects the anisotropy at the B site, with Bi causing a larger anisotropy than La.

\subsection{$a^0a^0c^+$ and $a^0a^0c^-$ structures}
In this section, we consider the effect of a single in-phase AFD ($a^0a^0c^+$) and out-of-phase AFD ($a^0a^0c^-$) on the SIA energy.
Since the symmetry is tetragonal, we use the following general expression to fit our SIA energy\cite{skomski2008}:
\begin{align}
 E_{SIA}(\theta,\phi) & = & K_1 \sin^2(\theta) + K_1' \sin^2(\theta) \cos(2\phi)\\
                      &   & + K_2 \sin^4(\theta) +   K_2'' \sin^4(\theta) \cos(4\phi)
 \label{eq:sia_general}
\end{align}
where $\theta\in[0,\pi]$ is the polar angle between the spin direction and the local $z$-axis and $\phi \in [0,2\pi]$ is the azimuthal angle in the plane perpendicular to $\theta$ = 0.

In Tab.\ref{tab:sia_aac} we report the values of the SIA constants for AFD rotations of 10$^\circ$ about the $z$ axis ($0^00^010^+$ and $0^00^010^-$) in both LFO and BFO.
In the two compounds we obtain $K_1>>K_2>K_2''>0$ for $a^0a^0c^+$, while for $a^0a^0c^-$ $K_1>0$ only for BFO.
According to Eq.\ref{eq:sia_general}, $K_1>0$ means that the anisotropy is predominantly uniaxial.
Because of the small size of $K_2$ and $K_2''$ relative to $K_1$, a second-order uniaxial model serves as a good description for the SIA.
It is interesting to see that the larger anisotropy of BFO versus LFO reported in the cubic structure is further emphasized in the presence of an oxygen octahedral rotation: $K_1$ is several hundred $\mu$eV for BFO while it is only few $\mu$eV for LFO.
The uniaxial anisotropy of BFO is very robust against the oxygen AFD distortions, however we remark that in LFO $a^0a^0c^+$ gives rise to uniaxial SIA ($K_1>0$) while $a^0a^0c^-$ gives rise to easy plane SIA ($K_1<0$).
This can be related to the small value of the SIA of LFO that can be easily affected by small structural changes.

\begin{table}[htbp!]
\begin{tabularx}{\columnwidth}{X X X X X r}
\hline
\hline
 &  & $K_1$ & $K_1'$  & $K_2$  & $K_2''$ \rule[-1ex]{0pt}{3.5ex}\\
\hline
$0^00^010^+$  & LFO  & 6.7   & 0 &  1.4 & 0.3  \rule[-1ex]{0pt}{3.5ex}\\
              & BFO & 264.0 & 0 & 3.5 & 0.7 \rule[-1ex]{0pt}{3.5ex}\\
$0^00^010^-$  & LFO & -1.6   & 0 & 1.3 & 0.2  \rule[-1ex]{0pt}{3.9ex}\\
              & BFO  & 235.3 & 0 & 4.3 & 0.8 \rule[-1ex]{0pt}{3.5ex}\\
\hline
\end{tabularx}
\caption{SIA constants ($\mu$eV) for LFO and BFO with $a^0a^0b^{+}$ and $a^0a^0b^{-}$ types of rotation (rotation amplitude of 10$^\circ$).}
\label{tab:sia_aac}
\end{table}

\begin{figure}[htbp!]
 \centering
 \includegraphics[width=7cm,keepaspectratio=true]{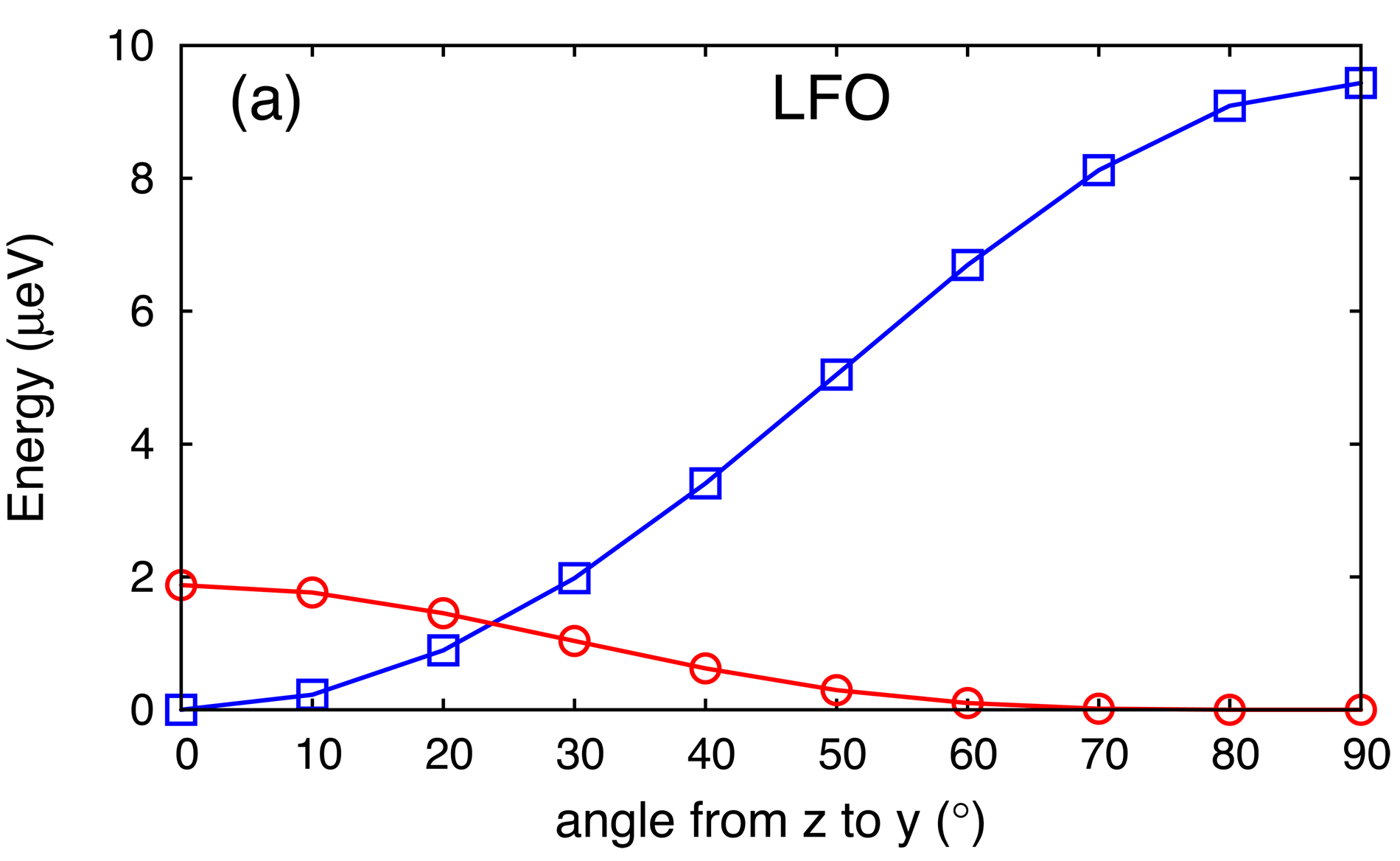}
 \includegraphics[width=7cm,keepaspectratio=true]{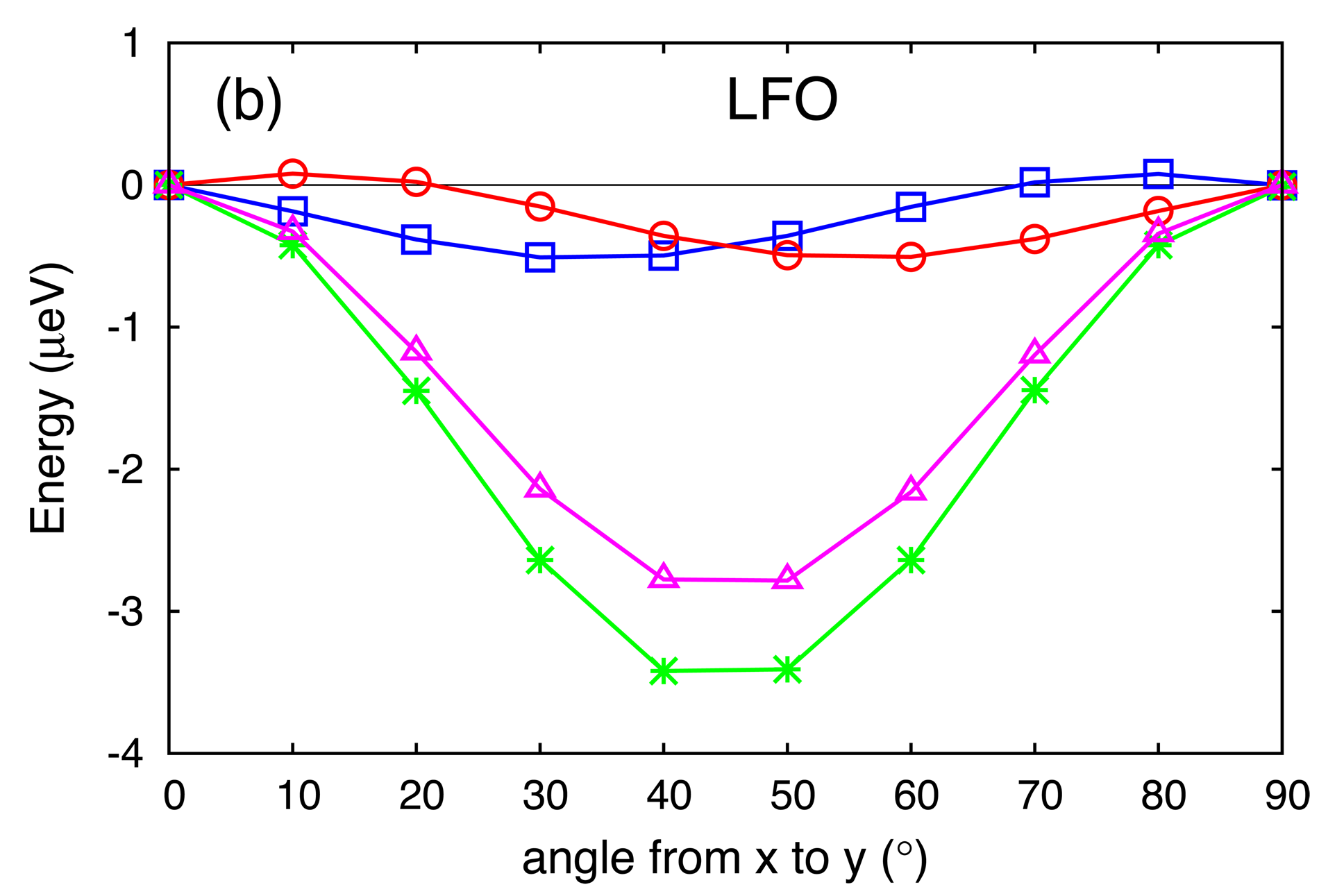}
 \includegraphics[width=7cm,keepaspectratio=true]{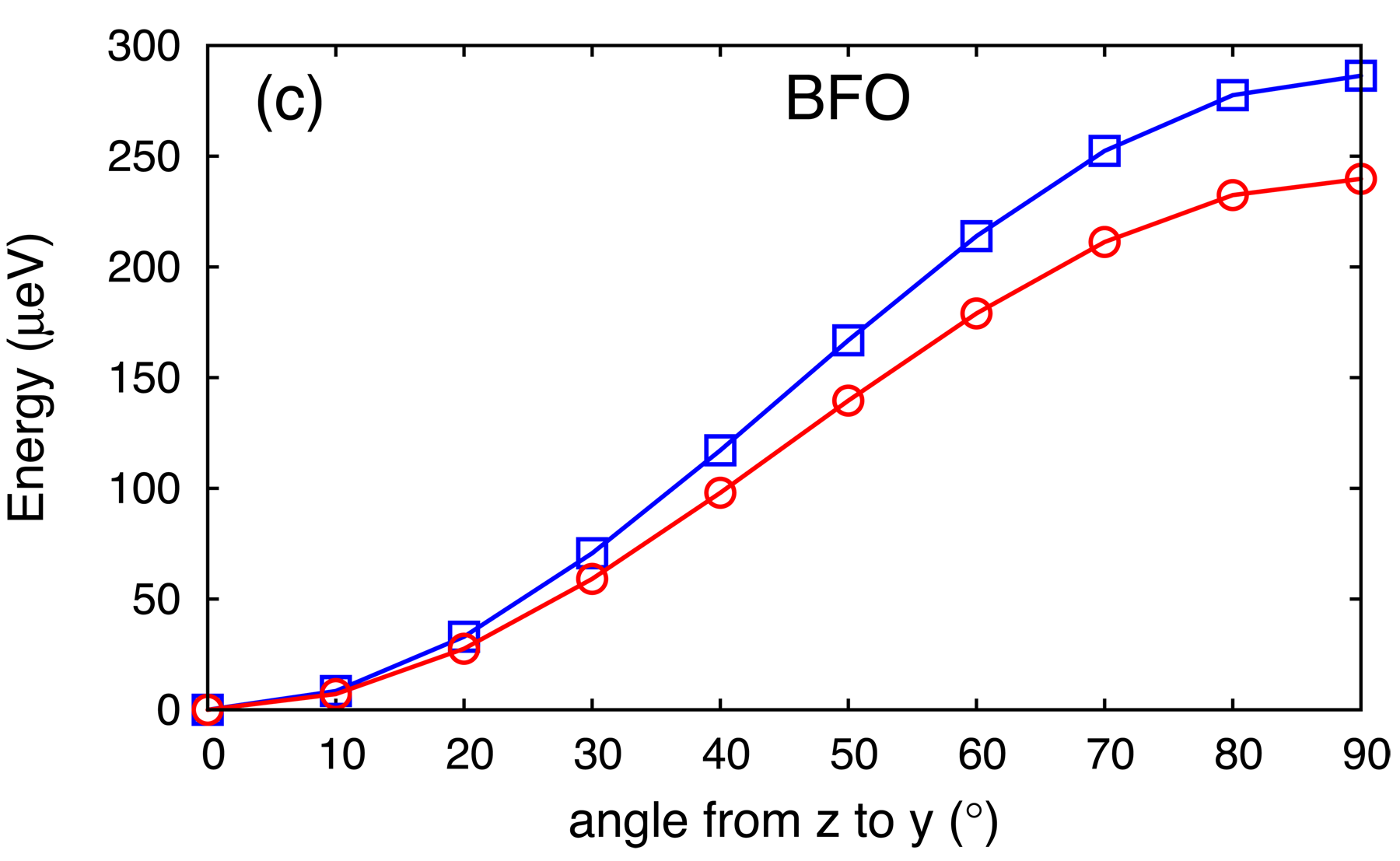}
\caption{SIA energy of LFO in (a) the $zy$-plane of LFO for $a^0a^0c^+$ (blue squares) and $a^0a^0c^-$ (red circles), (b) in the $xy$ plane in the presence of $a^0a^0c^+$ (blue squares for SIA at site A and red circles for site B, see text) and (c) SIA energy of BFO in the $zy$ plane in the presence of $a^0a^0c^+$ (blue squares) and $a^0a^0c^-$ (red circles).
In panel (b) we also report the total SIA energy (green stars) made of the sum of all the local SIA of each Fe sites.
We also compare in panel (b) the magnetocrystalline energy calculated by turning all the spins (no Al replacement) in the $xy$ plane with the same angle.
In panels (a) and (c), the angle 0$^\circ$ represents the [001] direction ($z$) and the angle 90$^\circ$ represents the [010] ($y$) direction.
In panel (b) the angle 0$^\circ$ represents the [100] direction and the 90$^\circ$ angle represents the [010] direction.
Note the energy amplitude differences between panels (a), (b) and (c).}
\label{fig:aac_sia}
\end{figure}

In Fig.\ref{fig:aac_sia}.a we plot the energy variation when the spin is turned in the $yz$ plane ($xz$ plane being equivalent) of LFO.
As we can see, the lowest energy is obtained for spins orientated along the $z$ axis, which means that when one AFD rotation develops in the perovskite structure, the rotational axis becomes the easy axis for the spins.
The SIA in the $yz$-plane (or $xz$-plane) with the presence of an AFD rotation along the $z$ axis is larger than that show for the perfect cubic cell but still stays small in LFO (6.7 $\mu$eV) while it is sizeable in BFO (264 $\mu$eV).

In Fig.\ref{fig:aac_sia}.b we also show the SIA energy of LFO in the plane perpendicular to the oxygen rotation axis ($xy$ plane).
One can see the spin does not point in the same direction on Fe atom site A (where the oxygen rotation is counter clockwise around Fe) and B (where the oxygen rotation is clockwise around Fe).
On atom site A the energy minimum is at 35$^\circ$ (blue squares on Fig.\ref{fig:aac_sia}.b) while for atom site B the energy minimum is at 55$^\circ$ (red diamonds on Fig.\ref{fig:aac_sia}.b).
This clearly show that the local spin anisotropy and thus the local spin direction follows directly the AFD rotation amplitude (45$^\circ\pm10^\circ$).
However, the global anisotropy is given by the sum of the SIA of all of the magnetic cation sites.
This sum of SIAs is represented in Fig.\ref{fig:aac_sia}.b by the purple triangles.
The global anisotropy gives rise to a minimum of energy at 45$^\circ$, highlighting the fact that the shift of $\pm$10$^\circ$ with respect to the [110] direction of each site compensate each other such that the global anisotropy is lowest in the [110] direction.
In Fig.\ref{fig:aac_sia}.b we also report the global SIA calculated by turning simultaneously the spins with all A and B site occupied with  Fe atoms (green dots).
With all the spins together, we recover the energy minimum at 45$^\circ$ and with energy amplitude in good agreement with the sum of the single spins result, highlighting the correctness of our approximation to replace all the surrounding Fe atoms by Al atoms.
It is also interesting to see that, apart from a phase shift of 10$^\circ$, in the plane $xy$ perpendicular to the AFD rotation axis, the amplitude of the anisotropy is similar to that found in the cubic structure.
We note that this global SIA is the magnetocrystalline anisotropy (MCA).
Comparing with BFO, the SIA energy in the $zy$ plane (Fig.\ref{fig:aac_sia}.c, red triangles) gives the same easy axis (the [001] direction minimizes the energy) but with a much larger amplitude (close to 300 $\mu$eV).

Performing the same SIA analysis with the $a^0a^0c^-$ structure and the same angle of 10$^\circ$ of oxygen octahedral rotation, we recover similar results as in the $a^0a^0c^+$ case but with some differences between BFO and LFO.
In BFO, the shape and amplitude of the SIA in $a^0a^0c^-$ is the same as in $a^0a^0c^+$, a result not surprising since locally, for one single spin, the first nearest-neighbors are the same in both $a^0a^0c^-$ and $a^0a^0c^+$ structures: In the $xy$-plane we have the same alternating A and B sites while in the $z$ direction the $A$ and $B$ sites, are inverted in $a^0a^0c^-$ (out-of-phase rotation along the $z$ direction) with respect to $a^0a^0c^+$ (in-phase rotation along the $z$ direction).
However this does not hold for LFO since the easy axis changes from [001] in $a^0a^0c^+$ to [111] in $a^0a^0c^-$.
This can be related to the fact that the anisotropy is much smaller in the case of LFO (few $\mu$eV) and so small changes in the structure, which give rise to changes in the SIA of a few $\mu$eV, can change the easy axis.

\subsection{$a^{0}b^{+}b^{+}$ and $a^{0}b^{-}b^{-}$ structures}
In this section we analyze the effect of two in-phase and two out-of-phase AFD rotations ($a^0b^+b^+$ and $a^0b^-b^-$) on the SIA.
We consider angles of AFD rotation of the same amplitude ($0^010^+10^+$ and $0^010^-10^-$) and we compare  with the previous case where we considered only one rotation.
The structures of $a^0b^+b^+$ and $a^0b^-b^-$ have space groups $I4/m$ and $Imma$ respectively and the model that describes the SIA is that given by Eq.\ref{eq:sia_general}.

In Tab.\ref{tab:sia_acc} we report the fitted $K$ parameters obtained in the $0^010^{+}10^{+}$ and $0^010^{-}10^{-}$ structure. 
The quantization axis ($\theta=\phi=0$ in Eq.\ref{eq:sia_general}) is the local hard axis, that is the [011] direction for all the cases reported in Tab.\ref{tab:sia_acc}.
In both compounds we obtain an easy plane SIA ($K_1$ is negative). 
As in the $a^0a^0b^{+/-}$ cases, the second-order anisotropy constants $K_1$ and $K_1'$ are larger than the fourth-order $K_2$ and $K_2''$. 
Again, we find the anisotropy of BFO (474.7 $\mu$eV) to be much higher than that of LFO (130.2 $\mu$eV) for both, out-of-phase and in-phase AFD distortions.
Interestingly, and in contrast to the case with oxygen rotation in only one direction, the SIA is strongly non-uniaxial since it is modulated by the large value of the $K_1'$ constant.
We also remark that the SIA energies increase when adding a second AFD distortion to the $a^0a^0b^{+/-}$ structures.

\begin{table}[htpb!]
\centering
\begin{tabularx}{\columnwidth}{X X X X X r}
\hline
\hline
 &  & $K_1$ & $K_1'$  & $K_2$  & $K_2''$ \rule[-1ex]{0pt}{3.5ex}\\
\hline
$0^010^+10^+$ & LFO & -71.8  & 24.1  & -1.6 & 0   \rule[-1ex]{0pt}{3.5ex}\\
              & BFO &  -515.6& 201.5 & -2.9 & 0 \rule[-1ex]{0pt}{3.5ex}\\
\hline
$0^010^-10^-$ & LFO & -130.2  & 27.2 & -1.4 & 0  \rule[-1ex]{0pt}{3.9ex}\\
              & BFO & -474.7 & 204.9 & -2.7  & 0 \rule[-1ex]{0pt}{3.5ex}\\
\hline
\end{tabularx}
\caption{Fitted SIA constants ($\mu$eV) of LFO and BFO in $a^0b^+b^+$ and $a^0b^-b^-$ structures with $b$=10$^\circ$.}
\label{tab:sia_acc}
\end{table}

\subsection{\textit{$a^-b^+a^-$ $Pnma$ structure}}
As for the previously reported distorted perovskites, the SIA of the $Pnma$ phase is also described by Eq.\ref{eq:sia_general}.
Our calculated coefficients for LFO and BFO in their $Pnma$ phase are given in Tab.\ref{tab:sia_pnma}.
As obtained for the previous cases with one ($a^0a^0c^+$ and $a^0a^0c^-$) and two ($a^0b^+b^+$ and $a^0b^-b^-$) oxygen rotations, the $K_1$ constant is the largest.
However, and in contrast to these previous strucures, in the $Pnma$ phase $K_1$ is negative, which means that the shape of the SIA is mainly of an easy plane rather than an easy axis type.
Here again we remark that the anisotropy of BFO is much larger than that of LFO.
For LFO $K_1'$ is small (7$\mu$eV) while it is larger for BFO (32 $\mu$eV), showing that the SIA for both LFO and BFO slightly deviate from a perfect easy plane.
The $K_2''$ and $K_2$ are zero for BFO and only $K_2''$ is zero for LFO, with $K_2$ having a small negative value.

\begin{table}[htpb!]
\centering
\begin{tabularx}{\columnwidth}{X X X X X r}
\hline
\hline
 &  & $K_1$ & $K_1'$  & $K_2$  & $K_2''$ \rule[-1ex]{0pt}{3.5ex}\\
\hline
$7^-8^+7^-$ & LFO 		& -158 & 7 & -6 & 0  \rule[-1ex]{0pt}{3.5ex}\\
 & BFO 				& -402 & 32 & 0 & 0    \rule[-1ex]{0pt}{3.5ex}\\
\hline
\end{tabularx}
\caption{Calculated SIA constants ($\mu$eV) fitted to Eq.\ref{eq:sia_general} for the $Pnma$ phase of LaFeO$_3$ and BiFeO$_3$.}
\label{tab:sia_pnma}
\end{table}

\begin{figure}[htbp!]
\centering
\includegraphics[width=7cm,keepaspectratio=true]{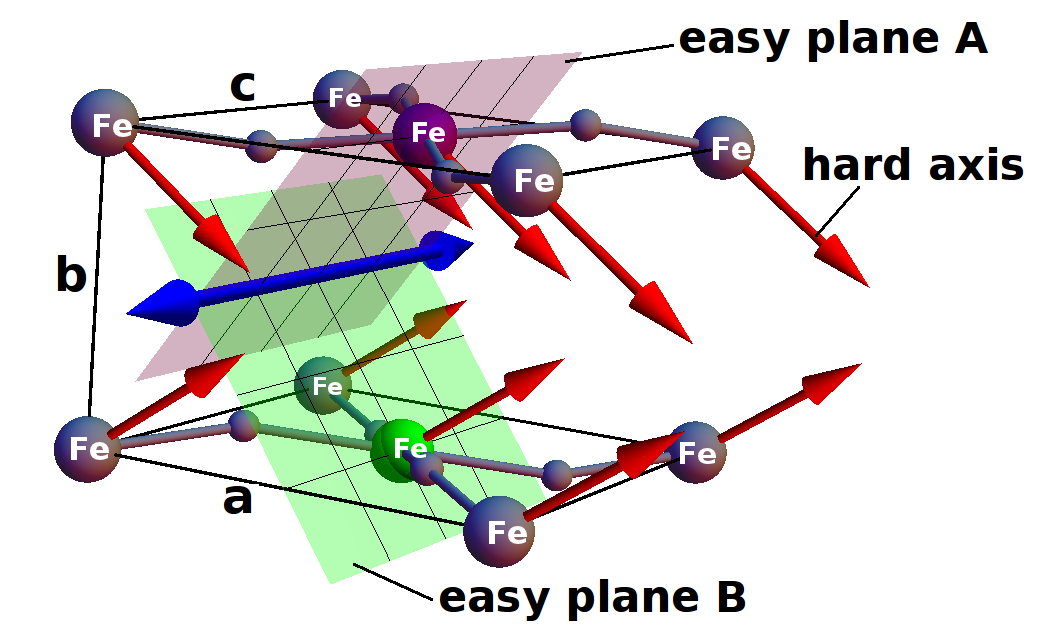}
\caption{Local SIA hard axis (red arrows) of two successive ac-planes of $Pnma$ structure (LFO). 
The intersection of two succesive easy planes along the $b$ direction (green and pink planes) is indicated by the blue arrow, showing that the two different easy planes give rise to an easy axis along the $c$ direction.
The hard axes point along the [1,1,0] and [1,-1,0] directions for the spins in plane B and A respectively.}
\label{fig:pnma_sia}
\end{figure}

In Fig.\ref{fig:pnma_sia} we show a schematic view of the direction of the hard axis (red arrows) and easy planes of the $7^-8^+7^-$ structure of LFO.
Each local hard axis points in the direction corresponding to the diagonal of the $ab$-planes in orthorhombic coordinates ($\theta$=54.7$^\circ$, $\phi$=45$^\circ$).
Interestingly, the hard axis of all Fe sites in successive $ac$-planes point alternately in opposite directions along the $b$ direction, as clearly shown in Fig.\ref{fig:pnma_sia} by the red arrows.
We then have two possible easy planes (noted A and B on Fig.\ref{fig:pnma_sia}) depending of the position of the Fe along the $b$ axis).
If we look at the combination of the easy planes A and B, we find that their intersection is aligned along the $c$ axis.
It is clear that if we introduce the strong Heisenberg exchange ($J\sim$ 6 meV) that has the tendency to align the spins  antiparallel, then the intersection of the easy planes will determine the spin direction as reported previously for $Pnma$ perovskites \cite{treves1962, peterlinneumaier1986}.
However, while it has been shown from symmetry arguments by Bertaut \cite{bertaut1963} that the SIA allows a canting of the spins in this system, here we prove that the combination of all the SIAs compensate in such a way that they do not give rise to a canting of the spins.

\subsection{SIA in $a^-a^-a^-$ $R\overline{3}c$ and $R3c$ structures}
In Tab.\ref{tab:sia_r3c} we report the SIA parameters obtained for BFO and LFO in the $R\bar3c$ ($a^-a^-a^-$) structure and in the $R3c$ structure that includes FE distortions in addition to the $a^-a^-a^-$ AFD distortions.
As for the $Pnma$ phase, in the $R\bar3c$ phase both LFO and BFO have large and negative $K_1$ constant (-168 $\mu$eV for LFO and -400 $\mu$eV for BFO) showing that the SIA is of easy plane form.
Since the other parameters are equal to zero ($K_1'$ and $K_2''$), we can consider that the SIA in $R\bar3c$ is purely easy plane anisotropy.
The orientation of this easy plane is perpendicular to the hard axis [111].
The spins can thus lie freely with any orientation in this easy plane which corresponds to the plane of rotation of the experimentally observed spin spiral structure of bulk BFO \cite{sosnowska1982}.

\begin{table}[htbp!]
\centering
\begin{tabularx}{\columnwidth}{X X c c c r}
\hline
\hline
                  &     & $K_1$ & $K_1'$  & $K_2$ & $K_2''$ \rule[-1ex]{0pt}{3.5ex}\\
\hline
$9^-9^-9^-$       & LFO & -168 	& 0 	& -1.6 & 0 \rule[-1ex]{0pt}{3.5ex}\\
$9^-9^-9^-$       & BFO & -400	& 0 	& -3.4 & 0  \rule[-1ex]{0pt}{3.5ex}\\
$9^-9^-9^-$+0.5FE & BFO & -281	& 0	& - & 0 \rule[-1ex]{0pt}{4.1ex}\\
$9^-9^-9^-$+1.0FE & BFO & -1.3	& 0	& - & 0 \rule[-1ex]{0pt}{3.5ex}\\
$9^-9^-9^-$+1.5FE & BFO & 139	& 0	& - & 0 \rule[-1ex]{0pt}{3.5ex}\\
$9^-9^-9^-$+1.0FE & LFO & -58	& 0  	& - & 0  \rule[-1ex]{0pt}{3.5ex}\\
$a^0a^0a^0$+1.0FE & BFO & 217	& 0	& -1.9 & 0 \rule[-1ex]{0pt}{4.1ex}\\
\hline
\end{tabularx}
\caption{Calculated SIA constants ($\mu$eV) obtained by fitting to Eq.\ref{eq:sia_general} of the $R\bar3c$ ($9^-9^-9^-$) and $R3c$ ($9^-9^-9^-$+FE) phases of LFO and BFO.
The factor in front of FE describes the amplitude of the FE distortions along the [111] direction such that the 1.0 amplitude is the point where the energy is minimum.
For comparison, the last line also shows the SIA constants with only the FE distortions ($R3m$ phase), \emph{i.e.} by removing the AFD distortions.\label{tab:sia_r3c}}
\end{table}

Adding a ferroelectric distortion along the diagonal ($R3c$ phase) changes completely the SIA of BFO.
The amplitude of the anisotropy has the tendency to evolve from easy plane ($K_1<0$) to easy axis ($K_1>0$) as the ferroelectric distortions increase.
It is interesting to see that at the point where the energy is minimum (1.0FE in Tab.\ref{tab:sia_r3c}) the SIA is close to the transition where $K_1$ changes sign and thus the anisotropy energy is strongly reduced.
To separate the effect of the FE distortion on the SIA, we also performed calculations in the presence of the FE distortion alone ($R3m$ phase in Tab.\ref{tab:sia_r3c}).
The SIA is then of easy axis form and it has a large value, close to the value of the $R\bar3c$ case ($9^-9^-9^-$).
Therefore, the SIAs induced by FE and $a^-a^-a^-$ distortions are in competition in the $R3c$ phase, since $a^-a^-a^-$  favors easy plane with $K_1<0$ and FE favors easy axis with $K_1>0$.
The resulting SIA in $R3c$ is then determined by the relative amplitudes of the FE and AFD distortions.
Replacing Bi by La in the same structure ($9^-9^-9^-$+1.0FE), we also find a reduction of the anisotropy but with a less strong effect.

The origin of the differences between La and Bi in the SIA in the compounds is difficult to understand.
In principle, it could come from the difference of atomic number, of cation size or because of the presence of a lone pair on the Bi atom.
To attempt to isolate these effects, we performed ``computer experiments'' by looking at the effect of the $A$-sites on the SIA. We report these in the next section.

\subsection{Differences in SIA between LFO and BFO}
For all the structures for which we calculated the SIA, we found the anisotropy to be larger in BFO than in LFO by up to a factor of 40.
This cannot be attributed to a volume effect or to different distortion amplitudes since in all cases we adopted the same cell parameters and the same amplitude of oxygen rotations or FE distortions.
This change in the anisotropy can then only originate from the differences between Bi and La cations.
A first difference between Bi$^{3+}$ and La$^{3+}$ that we can point to is the presence of the 6$s^2$ lone pair on Bi. 
In principle, this could affect the SIA of the magnetic Fe cation since an $A$-site lone pair has the tendency to modify the character of the bondlengnth of the $B$-site cations and anions\cite{seshadri2001,seshadri2006}.
Another important difference is the atomic number (Z) which is larger for Bi (83) than for La (57). 
The ionic radii, however, are roughly the same for Bi$^{3+}$ and La$^{3+}$ (around 117 pm).

In order to understand the $A$-site effect on the SIA, we performed computer experiments in which we replaced the $A$-site in $A$FeO$_3$ by different atoms having a lone pair or not and having different atomic numbers and ionic radii.
For a lone pair candidate we performed calculations with Sb at the $A$-site which has an atomic number (51) close to that of La but a smaller ionic radius (90 pm for Sb$^{3+}$).
For non lone pair candidates, we chose Tl which has an atomic number (81) close to that of Bi and a small radius (89 pm), and Y which has a much smaller atomic number (39) but an intermediate radius (104 pm).
We report the value of the $K_1$ constant for these different $A$ sites and for different structures in Tab.\ref{tab:sia_A-site}.

\begin{widetext}
 
\begin{table}[htbp!]
\centering
\begin{tabularx}{\textwidth}{X X X X X X c}
\hline
\hline
$A$ cation  & $Z$ & radius  & lone pair & \multicolumn{3}{c}{  K$_1$ } \rule[-1ex]{0pt}{3.5ex}\\
          &  &              &           & $a^0a^0c^+$  & $a^-a^-a^-$ &  $a^-a^-a^-$+FE \rule[-1ex]{0pt}{3.5ex}\\
\hline
Y  & 39  & 104 & no  & 16  & -73  & -23 \rule[-1ex]{0pt}{3.5ex}\\
Sb & 51 & 90  & yes & 110 & -161  & 12 \rule[-1ex]{0pt}{3.5ex}\\
La & 57 & 117 & no  & 7   & -168 & -58 \rule[-1ex]{0pt}{3.5ex}\\
Tl & 81 & 89  & no  & 114 & -335 & -57 \rule[-1ex]{0pt}{3.5ex}\\
Bi & 83 & 117 & yes & 264 & -400 & -1 \rule[-1ex]{0pt}{3.5ex}\\
\hline
\end{tabularx}
\caption{$A$ atomic number (Z), radius of the $A^{3+}$ cation (pm) and SIA constant K$_1$ ($\mu$eV) from Eq.\ref{eq:sia_general} of $A$FeO$_3$ with $A$ = Bi, La, Tl, Sb and Y within the same frozen structures ($a^0a^0c^+$,c=10$^\circ$; $a^-a^-a^-$, a=9$^\circ$; $a^-a^-a^-$+1.0FE, a=9$^\circ$).
The column ''lone pair`` states wether the $A^{3+}$ ion has a $s$ lone pair.
}
\label{tab:sia_A-site}
\end{table}
\end{widetext}

Unfortunately, no obvious trend emerge.
In the $a^0a^0c^+$ structure, all $A$-sites have $z$-easy axis SIA. 
Tl and Sb show similar anisotropy energies, 114 $\mu$eV and 110 $\mu$eV respectively while Y shows a very small anisotropy energy of 16 $\mu$eV. 
However, for $a^-a^-a^-$ there is no obvious trend in SIA with atomic number, ionic radius or presence or absence of lone pair, and the origin of the effect of the $A$-site on the SIA is not obvious.
We remark that for $a^-a^-a^-$ they are yet all easy plane

If the lone pair has an effect on the SIA, we expect it to be more pronounced in the case where the FE distortions are present since the lone pair is stereochemically active in the FE phase.\cite{cohen1992, watson1999, seshadri2001, seshadri2006}.
By comparing the last two columns in Tab.\ref{tab:sia_A-site}, we find indeed that the FE distortion causes the largest changes in SIA  for compounds with a lone pair on the $A$-site (SbFeO$_3$ and BiFeO$_3$). 
As expected from the conclusions in the previous sections, the FE distortions clearly decrease the easy plane SIA energy of the $R\bar3c$ structure for all the compounds. 
The large SIA energy for BiFeO$_3$ in the $R\bar3c$ structure (-400 $\mu$eV) is decreased drastically by the FE distortion (-1 $\mu$eV). For SbFeO$_3$ the anisotropy is even changed from easy plane (-161 $\mu$eV) to easy axis (12 $\mu$eV).
Note that for consistency in this work we use a $U$ value of 5 eV on Fe for all $A$-site cations.
Changes in $U$ value with $A$-site and subsequent effects on the SIA will be the subject of future works.

\section{Discussion}
In the previous sections we have determined the amplitudes of the different magnetic interactions, $i.e.$ exchange, SIA and DM interactions.
In the $R\bar3c$ and $R3c$ phases, a canting of the spins is allowed by symmetry, giving rise to wFM.
It has also been demonstrated that the only possible mechanism that leads to wFM in $R\bar3c$ and $R3c$ is the antisymmetric DM coupling \cite{bertaut1963}.
Our finding that the calculated SIA gives rise to an easy plane perpendicular to the [111] direction and the DM vector has the tendency to cant the spins in this easy plane is consistent with these earlier symmetry analyses .
We can conclude that in $R\bar3c$ and $R3c$, the SIA is neither cooperative with nor competing against the DM vector since there is almost no SIA energy cost to cant a spin in the plane perpendicular to the [111] direction by the DM interaction.
The direction of canting being the same for all the magnetic cations (global wFM), the DM force then competes only with the exchange interaction which prefers an antiferromagnetic alignment.
Thus, the amplitude of the wFM is directly related to the ratio between DM and $J$ such that larger DM and smaller $J$ favor larger wFM.

In the $Pnma$ structures, symmetry analysis shows that both DM and SIA allow for wFM \cite{bertaut1963}.
Our calculations showed that, even if it is allowed by symmetry, the SIA does not give rise to wFM.
While the local SIA of different spin sublattices are not the same, the sum of the SIAs on all the spin sublattices gives rise to a fixed unidirectional orientation of the spins (global easy axis or MCA).
As we saw in Fig.\ref{fig:pnma_sia}, the local easy planes are not parallel to the direction where the DM forces want to cant the spins.
This means that in the $Pnma$ structure, the canting of the spins is, as in the $R\bar3c$ structure, only due to the DM interaction. 
In contrast to the $R\bar3c$ case, where the SIA does not affect the development of the wFM, in the $Pnma$ structure the SIA competes with the DM.
As a result, in the $Pnma$ phase, the DM has to compete with not only the exchange interaction but also the SIA.

Our results allow us to conclude that depending on the structure, the amplitude of the canting of the spins in $Pnma$ and $R\bar3c$ phases comes from a delicate balance between SIA, DM and $J$.
If one wants to design large wFM through magnetic interactions in these distorted perovskites, one has to reduce $J$ and SIA relative to the DM interaction.
The main parameters which we can play with to design large canting are the amplitude of the FE and AFD distortions of the perovskite structure.
To that end, we need to understand the coupling between the crystal distortions and the canting of the spins, a link we analyze in the next section.

\section{Effect of AFD amplitude on $J$, DM and SIA}
As seen in the previous sections, it is clear that AFDs have the tendency to reduce the exchange parameter $J$, to induce large SIA and to allow for DM in the perovskite structure. 
As a result, they give rise to wFM in $Pnma$ and $R\bar3c$ phases for example.
In our calculations above, however, we calculated the amplitude of the wFM for only one amplitude of AFD distortions in each crystal phase. Here, we investigate how the amplitude of the crystal distortions acts on the non-collinear magnetism and thus on the wFM.
We thus analyze the influence of the distortions on the non-collinear magnetism in the $Pnma$ and $R\bar3c$ phases by calculating the evolution of $J$, DM, SIA and wFM as a function of the amplitude of AFD and FE distortions.

\begin{figure}[htbp!]
\includegraphics[width=7cm,keepaspectratio=true]{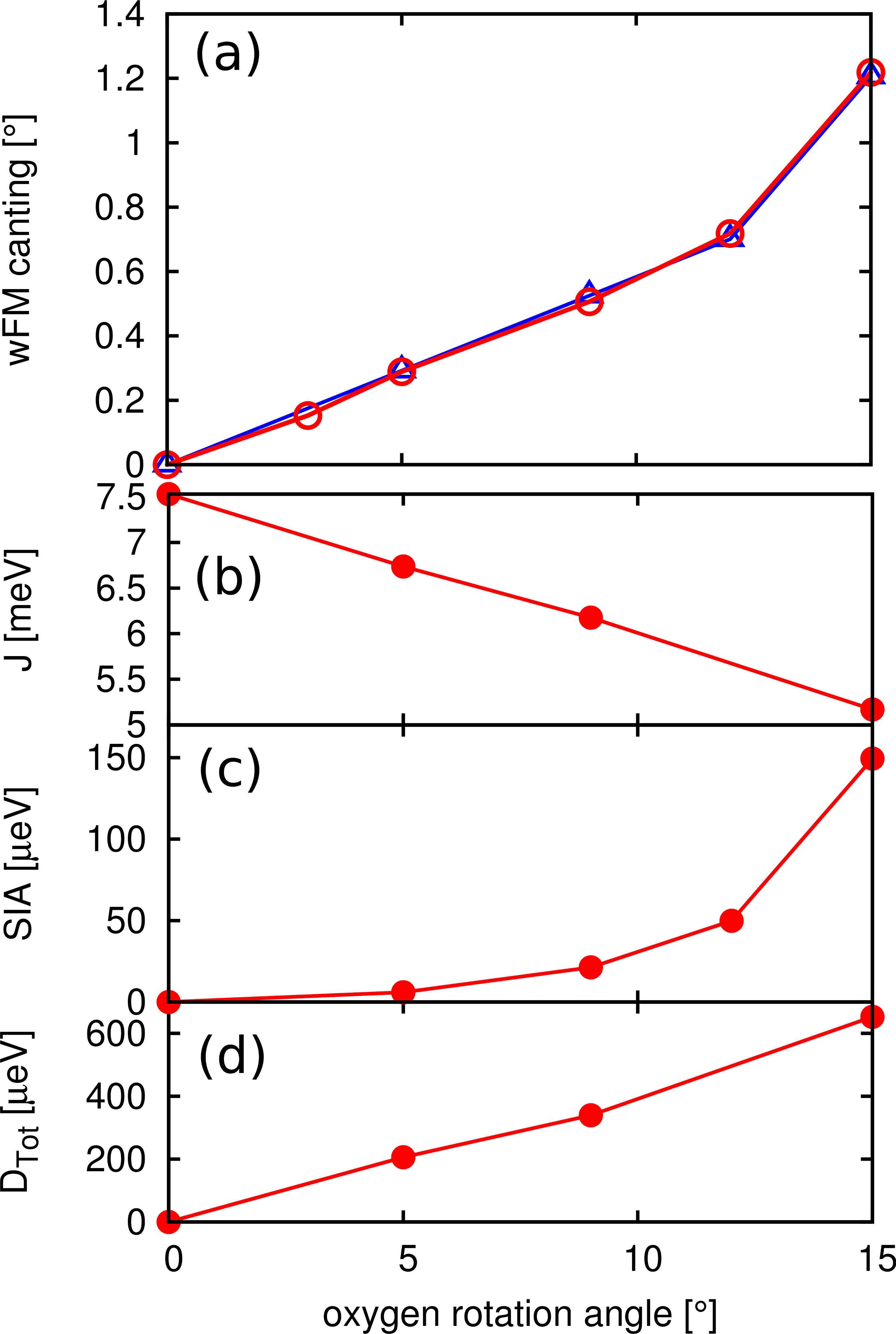} \\
\caption{Evolution of (a) the canting angle responsible for the wFM (red circles) (b) the exchange parameter $J$, (c) the SIA energy difference between hard and esasy direction of the spin and (d) the amplitude of the DM vector versus the oxygen rotation angle in $R\bar3c$-structure LFO.
In plot (a) we also report the values of $\arctan(D_{Tot}/J_{Tot})$ (blue triangles).}
\label{fig:wFMvsAFD_R-3c}
\end{figure}

In Fig.\ref{fig:wFMvsAFD_R-3c} (a) we report the evolution of the spin canting angle responsible for the wFM as a function of the AFD amplitudes in LFO in the $R\bar3c$ phase. 
For 0$^\circ$ to about 10$^\circ$ of AFD distortion, the wFM moment increases linearly with increasing rotation (red curve).
In Fig.\ref{fig:wFMvsAFD_R-3c} (b), (c) and (d) we report the evolution of respectively the total exchange parameter $J$, the SIA energy between the hard and easy direction and the total DM vector of the $R\bar3c$ structure as a function of the oxygen octahedral angle of rotation. 
The total $J$ is just the sum over all six nearest-neighbor exchange constants and D$_{Tot}$ is the norm of the total DM vector.
As expected, $J$ decreases with the increase of the amplitude of the oxygen octahedra rotation, and we find that the decrease is linear for rotation angles from 0$^\circ$ to 15$^\circ$.
In the same range, the DM increases linearly with the rotation angle and the SIA increases quadratically.
In Fig.\ref{fig:wFMvsAFD_R-3c} (a) we also report the value of $\arctan(D/J)$ which Interestingly follows perfectly the amplitude of the wFM.
This means that, when the SIA does not compete with DM, the amplitude of the spin canting can be directly related to the ratio between DM and $J$ and thus to the oxygen rotation amplitude.

\begin{figure}[htbp!]
\includegraphics[width=7cm,keepaspectratio=true]{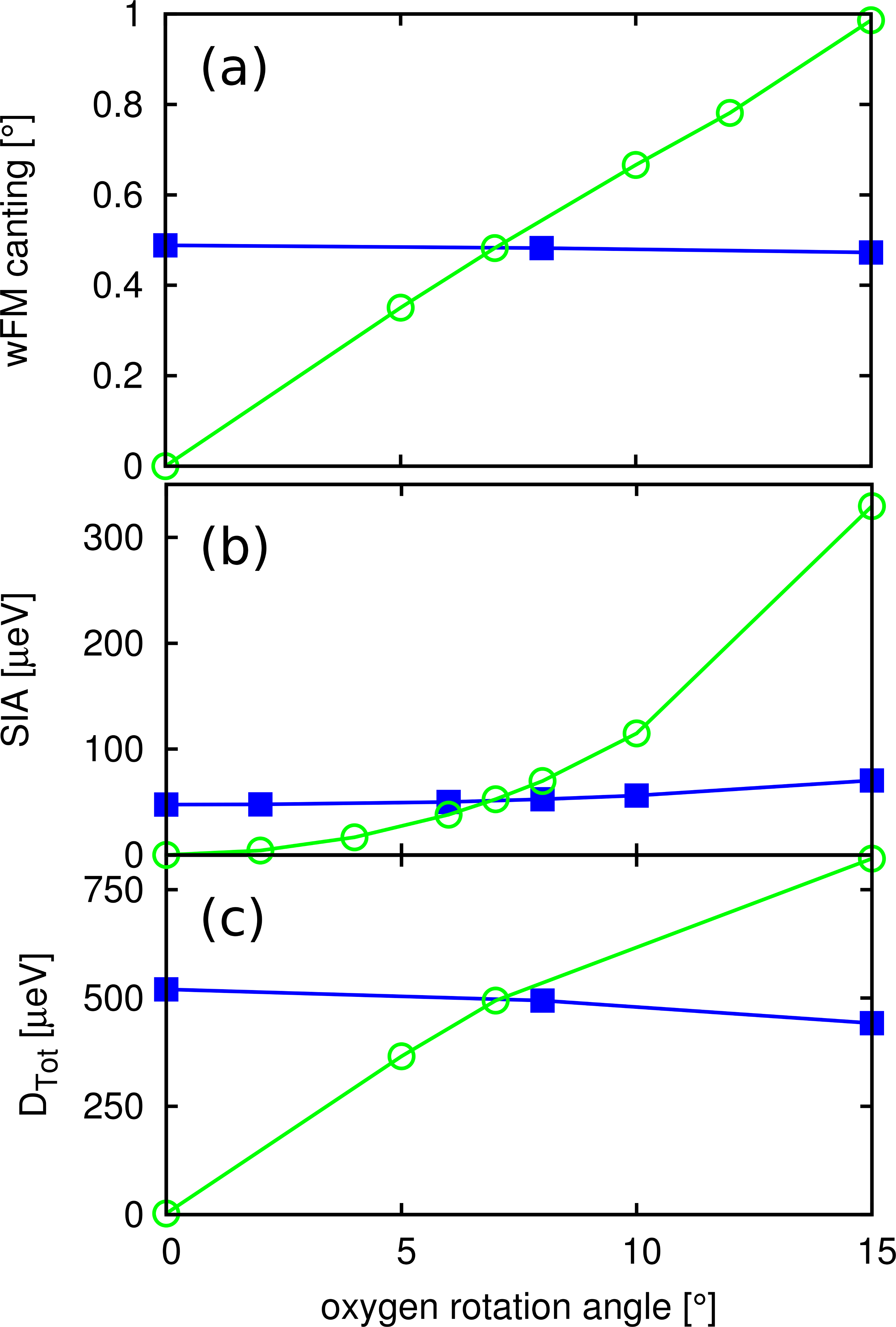} \\
\caption{Evolution of (a) the canting angle responsible for the wFM (b) the SIA energy difference between hard and esasy direction of the spin and (c) the amplitude of the DM vector versus the oxygen rotation angle of the in-phase rotation (blue squares) and the two out-of-phase rotations (green circles) of the $a^-b^+a^-$ structure of LFO. 
When the out-of-phase angle is changes, the in-phase angle is fixed at 8$^\circ$ ($x^-8^+x^-$) and when the in-phase rotation is varied, the out-of-phase rotations are fixed at 7$^\circ$ ($7^-x^+7^-$).
}
\label{fig:wFMvsAFD_Pnma}
\end{figure}

In Fig.\ref{fig:wFMvsAFD_Pnma} (a) we report the evolution of the spin canting angle responsible for the wFM versus the AFD amplitudes of LFO in the $Pnma$ phase.
Fixing the out-of-phase rotations and varying only the amplitude of the in-phase rotation causes almost no change in the amplitude of the wFM (blue squares in Fig.\ref{fig:wFMvsAFD_Pnma} (a)).
However, freezing the in-phase rotations and varying the out-of-phase rotations changes the magnitude of the wFM, clearly showing that the wFM is directly linked to the amplitude of the out-of-phase rotations (green circles in Fig.\ref{fig:wFMvsAFD_Pnma} a).
For the whole range of out-of-phase rotation angles considered (from 0$^\circ$ to 15$^\circ$), we observe that the relationship between AFD and wFM is linear while it deviates from perfect linear behaviour in the $R\bar3c$ structure beyond oxygen rotation angle of about 12$^\circ$ (Fig.\ref{fig:wFMvsAFD_R-3c} (a)).

In Fig.\ref{fig:wFMvsAFD_Pnma} (b) and (c) we report the evolution of SIA and the component of the DM vector that is responsible for wFM (the $x$-component as reported in section.\ref{sec:DM-pnma}).
As can be seen from the blue squares in Fig.\ref{fig:wFMvsAFD_Pnma} (b) and (c), the in-phase AFD has almost no effect on SIA and the DM interaction.
Only the out-of-phase oxygen rotations have a sizeable effect on the SIA and the DM interaction where, as for $R\bar3c$, SIA and $D_{Tot}$ increase with increasing out-of-phase AFD distortions.
This suggests that, to tune the wFM in the $Pnma$ type of structure, one has to modify the out-of-phase AFD distortions.

\section{Conclusion}
In this paper we analyzed the SIA, DM and exchange interactions in perovskite structures BiFeO$_3$ and LaFeO$_3$ with a range of structural distortions.
We analyzed the effect of these three interactions on the final magnetic ground state in different distortion patterns that can be present in the perovskites: $a^0a^0c^+$, $a^0a^0c^-$, $a^0b^+b^+$, $a^0b^-b^-$, $a^-b^+a^-$, $a^-a^-a^-$ and FE distortions.
We confirmed that in all cases the spin canting is due only to the DM interaction.
This is true even in the $a^-b^+a^-$ case where symmetry  allows also a possible contribution from the SIA \cite{bertaut1963}.
We found that the amplitude of the canting of the spins, and thus of the wFM, is determined by a balance between the amplitude of the DM interaction, the SIA and the exchange interaction.
The exchange interaction is always in competition with DM for the Fe$^{3+}$ of the studied compounds, while the SIA can be neutral, cooperative or in competition with DM.
For both chemistries we found that the SIA is not important in the non-ferroelectric $R\bar3c$ and $Pnma$ structures.
This is in contrast to the case of $Pnma$ LaMnO$_3$, where SIA seems to be the important mechanism that leads to wFM \cite{mozhegorov2007}.

We also looked at how the amplitude of the atomic distortions affect the different magnetic interactions and found that the change of the DM and the $J$'s are close to linear in the AFD amplitudes, while the SIA changes more drastically with these amplitudes.
The $J$ values are reduced by the AFD distortions while the DM vectors and SIA increase with the AFD distortions.
The amplitude of the wFM is then directly linked to the ratio between the amplitude of the DM interaction and the $J$ values such that the wFM canting angle follows a geometrical law with respect to this ratio and the angle of the oxygen rotations in the $R\bar3c$ and $Pnma$ cases.
It was also interesting to see that in the $Pnma$ case, mixing in-phase and out-of-phase AFD ($a^-b^+a^-$), the wFM is only induced by the out-of-phase AFD.
Thus, to control the wFM in the $Pnma$ type of structures, one has to tune the out-of-phase AFD.

By decomposing the magnetic interactions in BFO and LFO, we can conclude that the DM and $J$ parameters can be simply linked to the atomic distortions while the SIA has a much more complex behavior with respect to lattice modifications and is much more sensitive to them.
We note that the SIA values we computed here were all for the $d^5$ configuration of Fe$^{3+}$.
While a simple local atomic environment view of the SIA and crystal field splitting would lead us to predict small anisotropies ($L=0$ and thus small spin-orbit interaction) for $d^5$ electrons in ``cubic'' local oxygen octahedra (the Fe--O distances were kept constant and equal in our calculations), surprisingly we found for some structures very large anisotropies of the order of several hundred $\mu$eV.
This was mainly related to an $A$-site effect in the $A$BO$_3$ structure.
The size, the atomic number and the presence of a lone pair on the $A$-site seem to have a strong influence on the SIA and are responsible for the large anisotropies reported for the Fe$^{3+}$.
Since in our calculations we found a very small value of the orbital magnetic moment ($<$0.02 $\mu_B$), and since the the spin-orbit interaction term has the form  $\lambda L\cdot S$, the large induced SIA from the $A$ cation must be through the change in the spin-orbit coupling constant $\lambda$.
More complex interplay between the lattice distortions, the chemistry of the cations and the SIA can be expected when going away from $d^5$ electronic configuration of the $B$ cation since the SIA will also originate from a $L\neq0$ contribution in the spin-orbit interaction \cite{dai2008,mochizuki2009}.

Finally, we found that FE distortions, as present in the ground state of BFO, have a smaller effect on the DM and $J$ constants and thus on the wFM. 
However, we found that FE distortion has a strong influence on SIA since FE distortion favors an easy axis type of SIA which is in competition with the easy plane anisotropy induced by AFD distortions in the $R\bar3c$ structure.
This could be particularly interesting for technological applications for the following reason: 
the FE distortions tend to orient the spins along the [111] direction and the AFD distortions in the plane perpendicular to the [111] direction.
Then depending of the balance between the amplitude of the AFD and the amplitude of the FE (AFD/FE), the spins can lie either in the plane perpendicular to the [111] direction or parallel to the [111] direction.
When the spins lie in the [111] direction, no wFM is allowed by symmetry (because the DM vector is also along the [111] direction \cite{ederer2005a}).
Then, if one can simply tune the ratio AFD/FE in BFO by acting on the amplitude of these distortions (for example with pressure\cite{guennou2011}, epitaxial strain\cite{wang2003,ko2011,hatt2010}, electric field\cite{lebeugle2008} or phononic exitation), the wFM could be switched on and off through the competition between the SIA shapes (easy-plane versus easy-axis) of the AFD and FE distortions.
This process could be more readily attainable than the reversal of wFM that has been discussed previously\cite{fennie2008, benedek2011} and opens the door to new design rules for controlling magnetism by nonmagnetic external parameters.

\section{Acknowledgements}
We thank A. Scaramucci and M. Fechner for fruitful discussions.
This work was supported by the ETH Z\"{u}rich and FRS-FNRS Belgium (EB).

%

\end{document}